\documentclass[aps,prd,twocolumn,superscriptaddress,showkeys,floatfix]{revtex4-1}

\usepackage{graphicx}
\usepackage{amssymb}
\usepackage{amsmath}
\usepackage{bm}

\begin{document}


\title{Constraints on antisymmetric tensor fields from Bhabha scattering}



\author{Siddharth Tiwary}
\email{siddharth110200@gmail.com}
\affiliation{Department of Physics and Engineering Physics, 
University of Saskatchewan, 116 Science Place, Saskatoon, Canada SK S7N 5E2}
\affiliation{Department of Physics, Indian Institute of Technology Bombay,
Powai, Mumbai 400076, India}

\author{Rainer Dick}
\email{rainer.dick@usask.ca}
\affiliation{Department of Physics and Engineering Physics, 
University of Saskatchewan, 116 Science Place, Saskatoon, Canada SK S7N 5E2}



\begin{abstract} 
Antisymmetric tensor fields are a compelling prediction of string theory. 
This makes them an interesting target for particle
physics because antisymmetric tensors may couple to electromagnetic dipole
moments, thus opening a possible discovery
opportunity for string theory. The strongest constraints 
on electromagnetic dipole couplings would arise from couplings to electrons,
where these couplings would contribute to M\o{}ller and Bhabha scattering.
Previous measurements of Bhabha scattering constrain the couplings to 
$\tilde{M}_e m_C>7.1\times 10^4\,\mathrm{GeV}^2$, where $m_C$ is the mass of the 
antisymmetric tensor field and $\tilde{M}_e$ is an effective mass scale appearing in the 
electromagnetic dipole coupling. 
\end{abstract}


\maketitle



\section{Introduction\label{sec:intro}}

String theory has become an increasingly complex and compelling framework for
particle physics beyond the Standard Model. Discovering string signatures
through extra dimensions, $Z'$ bosons, Regge excitations, supersymmetry 
or string moduli has become a significant science driver for the development 
of next-generation accelerators \cite{ref:ILC,ref:CEPC}, and the advent 
of low-scale string models \cite{ref:IA} has increased the potential 
accessibility of all these signatures. 
While supersymmetry and Kaluza-Klein modes were traditionally studied as
possible indicators for the correctness of string theory for about half a 
century now, the prospects of low-scale string theory added the direct 
detection of string excitations as a further possibility 
\cite{ref:peskin,ref:bur1,ref:bur2,ref:LAhh,ref:chemtob,ref:LST,ref:dong,ref:kitazawa,ref:LAee,ref:AAregge}.
String signatures from Kaluza-Klein modes and $Z'$ bosons at the LHC have also 
been studied \cite{ref:LAKK,ref:AAZ}.

The discovery of Regge excitations would constitute a smoking gun signature for
string theory. However, the discovery of antisymmetric tensor fields would also
provide a very strong indication for the correctness of string theory.
Antisymmetric tensor excitations appear in the massless sector of closed 
strings and throughout the excited levels of closed and open strings, and
Kalb and Ramond have pointed out that antisymmetric tensor fields can also 
mediate gauge interactions between strings \cite{ref:KR}. Quantization of 
antisymmetric tensors leaves only a single externally propagating transverse
polarization state \cite{ref:notoph}. However, the couplings of the tensor are 
still restricted by Lorentz invariance and the non-physical transverse polarization 
states contribute as virtual states to scattering amplitudes. 
This makes antisymmetric tensor fields another interesting 
target for exploration of possible low-energy signals of string theory. 

Both standard string theory and the Kalb-Ramond proposal include
couplings of antisymmetric tensor fields $\mathcal{C}_{\mu\nu}(x)$ to string world 
sheets $\mathcal{S}$: $X^\mu(\tau,\sigma)\equiv X^\mu(\sigma^1,\sigma^2)$
in the form
\begin{eqnarray} \nonumber
S_{X\mathcal{C}}&=&\mu_s\int_{\mathcal{S}}\!\mathcal{C}
\\ \label{eq:SXC}
&=&\frac{\mu_s}{2}\int\!d^2\sigma\left(\dot{X}^\mu X'^\nu
-\dot{X}^\nu X'^\mu\right)\mathcal{C}_{\mu\nu}(X).
\end{eqnarray}
Since we are interested in low energy string phenomenology
we avoid the usual designation $B_{\mu\nu}$ for the antisymmetric tensor fields
to avoid confusion with the $U_Y(1)$ field strength in the Standard Model.
We also include a string charge $\mu_s$ with mass dimension 1 to have canonical
mass dimension 1 for the antisymmetric tensor field, such that the 
kinetic term for the Kalb-Ramond field strength $\mathcal{C}_{\mu\nu\rho}=
\partial_\mu\mathcal{C}_{\nu\rho}+\partial_\nu\mathcal{C}_{\rho\mu}+\partial_\rho\mathcal{C}_{\mu\nu}$
in four spacetime dimensions
can be written as $\mathcal{L}_{d\mathcal{C}}=-\,\mathcal{C}^{\mu\nu\rho}\mathcal{C}_{\mu\nu\rho}/6$.
The Kalb-Ramond picture of gauge interactions between strings also includes
dimensionless string boundary charges $g_s$ and a vector field $\mathcal{B}_\mu$
which couples to the boundary $\partial\mathcal{S}$ of open string world sheets,
\begin{equation}\label{eq:SXB}
S_{X\mathcal{B}}=g_s\int_{\partial\mathcal{S}}\!\mathcal{B}
=g_s\int\! d\tau\left[\dot{X}^\mu\mathcal{B}_{\mu}(X)\right]_{\sigma=0}^{\sigma=\ell}.
\end{equation}
This is appealing, because it yields a mass $m_C=\mu_s/\sqrt{2}g_s$ for the Kalb-Ramond field
in the four-dimensional spacetime action,
\begin{eqnarray}\nonumber
\mathcal{L}&=&-\,\frac{1}{6}\mathcal{C}^{\mu\nu\rho}\mathcal{C}_{\mu\nu\rho}
-\frac{1}{4}\mathcal{B}^{\mu\nu}\mathcal{B}_{\mu\nu}
+\frac{\mu_s}{2g_s}\mathcal{C}^{\mu\nu}\mathcal{B}_{\mu\nu}
\\ \label{eq:kinfields}
&&-\,\frac{\mu_s^2}{4g_s^2}\mathcal{C}^{\mu\nu}\mathcal{C}_{\mu\nu}
+j^{\mu\nu}\mathcal{C}_{\mu\nu}+j^\mu\mathcal{B}_\mu,
\end{eqnarray}
without breaking the KR gauge symmetries
\begin{eqnarray}\label{eq:KR1}
&&\mathcal{C}_{\mu\nu}\to \mathcal{C}_{\mu\nu}+\partial_\mu f_\nu-\partial_\nu f_\mu,
\\ \label{eq:KR2}
&&\mathcal{B}_\mu\to\mathcal{B}_\mu+(\mu_s/g_s)f_\mu+\partial_\mu f.
\end{eqnarray}
The string currents in (\ref{eq:kinfields}) are from (\ref{eq:SXC},\ref{eq:SXB})
\begin{eqnarray}\nonumber
j^{\mu\nu}(x)&=&\frac{\mu_s}{2}\int\! d\tau\!\int_0^\ell\! d\sigma
\left[\dot{X}^\mu(\tau,\sigma) X'^\nu(\tau,\sigma)
\right.
\\ \label{eq:stringcurrents1}
&&-\left.\dot{X}^\nu(\tau,\sigma) X'^\mu(\tau,\sigma)\right]\delta(x-X(\tau,\sigma)),
\end{eqnarray}
\begin{equation}\label{eq:stringcurrents2}
j^{\mu}(x)=g_s\!\int\! d\tau\left[\dot{X}^\mu(\tau,\sigma)
\delta(x-X(\tau,\sigma))\right]_{\sigma=0}^{\sigma=\ell}.
\end{equation}
They satisfy the consistency conditions
\begin{equation}
\partial_\mu j^{\mu\nu}(x)=(\mu_s/2g_s)j^\nu(x),\quad\partial_\mu j^{\mu}(x)=0.
\end{equation}
The gauge invariant field \cite{note1}
\begin{equation}
C_{\mu\nu}=\mathcal{C}_{\mu\nu}-\frac{g_s}{\mu_s}\mathcal{B}_{\mu\nu}
\end{equation}
satisfies the equations of motion
\begin{equation}\label{eq:sources1}
\partial_\mu C^{\mu\nu\rho}(x)-\frac{\mu_s^2}{2g_s^2} C^{\nu\rho}(x)=-\,j^{\nu\rho}(x)
\end{equation}
and
\begin{equation}\label{eq:sources2}
\partial_\mu C^{\mu\nu}(x)=\frac{g_s}{\mu_s}j^{\nu}(x).
\end{equation}
These equations imply that the Kalb-Ramond field in the interaction picture is
a transverse massive antisymmetric tensor field with mode expansion 
\begin{eqnarray} \nonumber
C_{\mu\nu}(x)&=&\int\!\frac{d^3\bm{k}}{4\sqrt{2\pi^3E(\bm{k})}}\,\epsilon_{\alpha\beta\gamma}
\epsilon^{(\beta)}_\mu(\bm{k})\epsilon^{(\gamma)}_\nu(\bm{k})
\\ \nonumber
&&\times[a^{(\alpha)}(\bm{k})\exp(\mathrm{i}k\cdot x)
\\ \label{eq:Cinter}
&&+\,a^{(\alpha)+}(\bm{k})\exp(-\,\mathrm{i}k\cdot x)].
\end{eqnarray}
We choose polarization vectors $\epsilon^{(\alpha)}_\mu(\bm{k})$
such that for $\alpha\in\{1,2\}$
\begin{equation}
k\cdot\epsilon^{(\alpha)}(\bm{k})=\bm{k}\cdot\bm{\epsilon}^{(\alpha)}(\bm{k})=0,
\end{equation}
whereas $\epsilon_0^{(3)}(\bm{k})\neq 0$,
\begin{equation}
k\cdot\epsilon^{(3)}(\bm{k})=0\neq\bm{k}\cdot\bm{\epsilon}^{(3)}(\bm{k}).
\end{equation}
Comparison with the Kalb-Ramond field in Coulomb gauge \cite{ref:shadow2} shows that 
the single completely transverse physical polarization state is given by
 $\epsilon^{(1)}_\mu(\bm{k})\epsilon^{(2)}_\nu(\bm{k})-\epsilon^{(2)}_\mu(\bm{k})\epsilon^{(1)}_\nu(\bm{k})$,
whereas the two (spatially longitudinal but 4d transverse) po\-lar\-iza\-tions
$[\bm{\epsilon}^{(\alpha)}(\bm{k})\otimes\bm{\epsilon}^{(3)}(\bm{k})
-\bm{\epsilon}^{(3)}(\bm{k})\otimes\bm{\epsilon}^{(\alpha)}(\bm{k})]_{\alpha\in\{1,2\}}$
are unphysical. Therefore only $a^{(3)+}(\bm{k})$ generates external physical states for
the Kalb-Ramond field, but the other transverse modes also contribute to virtual Kalb-Ramond
exchange. 

The normalization in (\ref{eq:Cinter}) was chosen such that the canonical commutation
relation
\begin{eqnarray} \nonumber
&&[C_{\mu\nu}(\bm{x},t),\partial_0 C_{\rho\sigma}(\bm{x}',t)]
=\mathrm{i}\delta^\perp_{\mu\nu\rho\sigma}(\bm{x}-\bm{x}')
\\ \nonumber
&&=\mathrm{i}\int\!\frac{d^3\bm{k}}{(2\pi)^3}\exp[\mathrm{i}\bm{k}\cdot(\bm{x}-\bm{x}')]
\\
&&\quad\times\frac{1}{2}
[P^\perp_{\mu\rho}(\bm{k})P^\perp_{\nu\sigma}(\bm{k})-P^\perp_{\mu\sigma}(\bm{k})P^\perp_{\nu\rho}(\bm{k})],
\end{eqnarray}
\begin{equation}
P^\perp_{\mu\rho}(\bm{k})=\epsilon^{(\alpha)}_\mu(\bm{k})\epsilon^{(\alpha)}_\rho(\bm{k}),
\end{equation}
yields
\begin{equation}
[a^{(\alpha)}(\bm{k}),a^{(\beta)+}(\bm{k}')]=\delta_{\alpha\beta}\delta(\bm{k}-\bm{k}').
\end{equation}
The propagator for the Kalb-Ramond field is 
\begin{eqnarray} \nonumber
&&G_{\mu\nu,\kappa\lambda}(x-x')=\mathrm{i}
\bm{\langle} 0\bm{|}\mathrm{T}\,C_{\mu\nu}(x)C_{\kappa\lambda}(x')\bm{|}0\bm{\rangle}
\\ \nonumber
&&=\int\!\frac{d^4k}{32\pi^4}\,
\frac{\exp[\mathrm{i}k\cdot(x-x')]}{k^2+m_C^2-\mathrm{i}\epsilon}
\\ 
&&\quad\times
[P^\perp_{\mu\kappa}(\bm{k})P^\perp_{\nu\lambda}(\bm{k})-P^\perp_{\mu\lambda}(\bm{k})P^\perp_{\nu\kappa}(\bm{k})].
\end{eqnarray}

This fits into canonical string theory if we assume that antisymmetric tensors
are spacetime manifestations of tensor excitations of strings. Another, more 
speculative interpretation of (\ref{eq:kinfields}) would suggest that strings
and quantum fields may co-exist, such that strings define a genuine
extension of quantum field theory without encompassing the quantum field theories
of particle physics as a mere low-energy effective description. Strings are then
classical objects (from the target space quantum field theory perspective) which carry
two-di\-men\-sion\-al quantum field theories for their embeddings, whereas point particles
are quantized in the standard way. In such a framework, antisymmetric tensors 
would act as mediators between point particles and strings.

Either way, antisymmetric tensor fields are unavoidable in string theory and 
we should study their possible signatures in particle physics experiments.

Antisymmetric tensors can couple in particular to electromagnetic dipole 
moments, thus contributing to M\o{}ller and Bhabha scattering. This is particularly 
relevant for upcoming or proposed lepton 
colliders \cite{ref:moller,ref:ILC,ref:CEPC,ref:FCCee}, because the well-defined
initial state in the scattering events facilitates the search for deviations from
Standard Model scattering cross sections. Existing data on Bhabha scattering from
previous $\mathrm{e}^+\mathrm{e}^-$ collider experiments already limit 
deviations from Standard Model cross sections, and here we report constraints on 
antisymmetric tensors using published data from 
TASSO \cite{ref:tasso,ref:tasso2}, PLUTO \cite{ref:pluto}, MAC \cite{ref:mac}, 
TOPAZ \cite{ref:topaz} and OPAL \cite{ref:opal}.

Although our primary interest is on collider-based constraints and prospects for
antisymmetric tensor fields as harbingers of string theory, we note in passing
that an antisymmetric tensor might also lend itself as a dark matter candidate due 
to its electroweak singlet properties. Within the coupling model (\ref{eq:LI}) that
we investigate for particle physics implications, this possibility is excluded from 
the requirement of longevity: Mass values in the MeV range or below for the antisymmetric 
tensor field are compatible with a lifetime of order $10^{18}$ s if the coupling to 
neutrinos satisfies $v_h/M_n\lesssim 10^{-16}$ (assuming $a_M^2+a_e^2\simeq 1$), 
i.e.~if the coupling scale $M_n$ is of order of the reduced Planck mass. 
However, light antisymmetric tensors with Standard Model couplings of
the form (\ref{eq:LI}) would have revealed their existence through resonances
in scattering experiments, whereas heavy antisymmetric tensors decay too fast 
to serve as dark matter. On the other hand, antisymmetric tensor fields could serve as 
messengers into a dark sector \cite{ref:adrd}, but we will focus on their corrections 
to M\o{}ller and Bhabha scattering in the following.

We discuss the contribution from antisymmetric tensors to M\o{}ller and Bhabha 
scattering in Sec.~\ref{sec:bhabha}. Constraints on the antisymmetric tensor
mass and the coupling to electrons are reported in Sec.~\ref{sec:results}.
Sec.~\ref{sec:conc} summarizes our conclusions.

\section{Bhabha scattering through Kalb-Ramond exchange}
\label{sec:bhabha}

The KR gauge invariant tensor $C_{\mu\nu}$ can have $SU_w(2)\times U_Y(1)$ invariant couplings
to Standard Model fermions through interaction terms
\begin{eqnarray}\nonumber
\mathcal{L}_I&=&-\,\frac{1}{M_e}\overline{\Psi}\cdot H S^{\mu\nu}(a_m+\mathrm{i}a_e\gamma_5)
\frac{1+\gamma_5}{\sqrt{2}}\psi_e C_{\mu\nu}
\\ \nonumber
&&-\,\frac{1}{M_n}\overline{\Psi}\cdot\tilde{H} S^{\mu\nu}(a_m+\mathrm{i}a_e\gamma_5)
\frac{1+\gamma_5}{\sqrt{2}}\psi_n C_{\mu\nu}
\\ \label{eq:LI}
&&+\,\mathrm{h.c.}
\end{eqnarray}
Here
\begin{equation}
\Psi=\left(\begin{array}{c}
\psi_n\\ \psi_e\\ \end{array}\right)
\end{equation}
are $SU_w(2)$ spinors with neutrino/up-type upper fields and electron/down-type lower fields
and 
\begin{equation}
\underline{H}=\left(\begin{array}{c}
H^+\\
H^0\\
\end{array}\right),\quad
\tilde{\underline{H}}=\underline{\epsilon}\cdot\underline{H}^*
=\left(\begin{array}{c}
\,\,\,\, H^{0,*}\\
-H^{+,*}\\
\end{array}\right)
\end{equation}
is the Higgs doublet.
We use the spinor representation of the Lorentz generators,
\begin{equation}
S^{\mu\nu}=\frac{1}{2}\sigma^{\mu\nu}=\frac{\mathrm{i}}{4}[\gamma^\mu,\gamma^\nu],
\end{equation}
in the dipole operators.
The leading order coupling to the Standard Model fermions from (\ref{eq:LI}) is then
\begin{equation} \label{eq:LeC}
\mathcal{L}_{eC}=-\,\frac{v_h}{M_e}\overline{\psi}_e S^{\mu\nu}(a_m+\mathrm{i}a_e\gamma_5)
\psi_e C_{\mu\nu},
\end{equation}
with the Higgs expectation value $v_h=\sqrt{2}\langle H^0\rangle$.

We assume unbroken KR gauge symmetry and therefore focus on the gauge invariant 
couplings (\ref{eq:LI}) of the KR fields to Standard Model fermions.
 Any mixing of a string boundary gauge field $\mathcal{B}_\mu$ with Standard Model
gauge fields, or promotion to a $Z'$ through a mass term, would break the KR gauge 
symmetry. These would then enhance the pool of string-motivated 
vector fields \cite{ref:luis,ref:AAZ} which could also contribute to $g-2$ values for 
Standard Model leptons \cite{ref:aa1,ref:aa2}. However, here we are content 
with the observation that breaking of KR gauge symmetry would provide further motivation 
and relevance for the study of massive vector fields from string theory. 

Absence of a resonance from antisymmetric tensor exchange in Bhabha scattering up to 
the highest LEP energies tells us that $m_C>209$ GeV. Furthermore, we have no reason
to expect a fundamental antisymmetric tensor field from string theory to be hadrophobic, 
and absence of Beyond the Standard Model resonances up to the highest energies probed in 
hadronic collisions indicates $m_C>1$ TeV \cite{ref:atlas1}. We will therefore analyze the 
coupling (\ref{eq:LeC}) under the assumption $\sqrt{s}\ll m_c$, since the highest collision 
energy used here is $\sqrt{s}=136.23$ GeV \cite{ref:opal}.

Exchange of virtual Kalb-Ramond tensor particles through the coupling (\ref{eq:LeC}) 
yields $t$ and $u$ channel contributions to M\o{}ller scattering and 
$s$ and $t$ channel contributions to Bhabha scattering. In order $\alpha_S v_h^2/M_e^2$, 
this shifts the corresponding cross sections at energy $\sqrt{s}$ 
by \cite{note2}
\begin{eqnarray}\nonumber
\left.\frac{d\sigma}{d\Omega}\right|_{v_h^2/M_e^{2}}&=&\frac{\pi^2}{4}s
\left(\mathcal{M}^{(\gamma,Z)+}\mathcal{M}^{(C)}
\right.
\\ \label{eq:dsigma1}
&&+\left.\mathcal{M}^{(C)+}\mathcal{M}^{(\gamma,Z)}\right),
\end{eqnarray}
where $\mathcal{M}^{(\gamma,Z)}$ are the M\o{}ller or Bhabha scattering amplitudes
through photon and $Z$ exchange and $\mathcal{M}^{(C)}$ are the corresponding amplitudes from
Kalb-Ramond exchange. Here we use a normalization of scattering amplitudes such
that the scattering matrix with incoming 4-momentum $P_i$ and final 4-momentum $P_f$ is
\begin{equation}
S_{fi}=-\,\mathrm{i}\mathcal{M}_{fi}\delta(P_f-P_i).
\end{equation}
We imply summation/averaging $(1/4)\sum_{s'_2,s'_s}\sum_{s_1,s_2}$ for the outgoing/incoming
spin orientations in the product of scattering amplitudes. Corrections for spin-po\-lar\-ized
M\o{}ller scattering will be of interest for the upcoming MOLLER experiment.

The amplitude for M\o{}ller scattering
 $\bm{|}\bm{k}_1,\sigma_1;\bm{k}_2,\sigma_2\bm{\rangle}\to\bm{|}\bm{p}_1,s_1;\bm{p}_2,s_2\bm{\rangle}$
through Kalb-Ramond exchange is then
\begin{eqnarray}\nonumber
&&\mathcal{M}^{(C)}_{--}=-\,\frac{v_h^2}{16\pi^2 M_e^2
\sqrt{E(\bm{p}_1)E(\bm{p}_2)E(\bm{k}_1)E(\bm{k}_2)}}
\\ \nonumber
&&\times\left(\frac{\overline{u}(\bm{p}_1,s_1)\Gamma^{\mu\nu}u(\bm{k}_1,\sigma_1)
\overline{u}(\bm{p}_2,s_2)\Gamma^{\kappa\lambda}u(\bm{k}_2,\sigma_2)}{
(k_1-p_1)^2+m_C^2-\mathrm{i}\epsilon}
\right.
\\ \nonumber
&&\times
\left[P^{\perp}_{\mu\kappa}(\bm{k})P^{\perp}_{\nu\lambda}(\bm{k})\right]_{\bm{k}=\bm{k}_1-\bm{p}_1}
-\left[P^{\perp}_{\mu\kappa}(\bm{k})P^{\perp}_{\nu\lambda}(\bm{k})\right]_{\bm{k}=\bm{k}_1-\bm{p}_2}
\\ \nonumber
&&\times
\left.\frac{\overline{u}(\bm{p}_2,s_2)\Gamma^{\mu\nu}u(\bm{k}_1,\sigma_1)
\overline{u}(\bm{p}_1,s_1)\Gamma^{\kappa\lambda}u(\bm{k}_2,\sigma_2)}{
(k_1-p_2)^2+m_C^2-\mathrm{i}\epsilon}
\right),
\end{eqnarray}
where
\begin{equation}
\Gamma^{\mu\nu}=S^{\mu\nu}(a_m+\mathrm{i}a_e\gamma_5).
\end{equation}

The amplitude for Bhabha scattering through Kalb-Ramond exchange is
\begin{eqnarray}\nonumber
&&\mathcal{M}^{(C)}_{-+}=-\,\frac{v_h^2}{
16\pi^2 M_e^2\sqrt{E(\bm{p}_1)E(\bm{p}_2)E(\bm{k}_1)E(\bm{k}_2)}}
\\ \nonumber
&&\times\left(
\frac{\overline{u}(\bm{p}_1,s_1)\Gamma^{\mu\nu} v(\bm{p}_2,s_2)
\overline{v}(\bm{k}_2,\sigma_2)\Gamma^{\kappa\lambda} u(\bm{k}_1,\sigma_1)}{
(k_1+k_2)^2+m_C^2-\mathrm{i}\epsilon}
\right.
\\ \nonumber
&&\times\left[P^{\perp}_{\mu\kappa}(\bm{k})P^{\perp}_{\nu\lambda}(\bm{k})\right]_{\bm{k}=\bm{k}_1+\bm{k}_2}
-\left[P^{\perp}_{\mu\kappa}(\bm{k})P^{\perp}_{\nu\lambda}(\bm{k})\right]_{\bm{k}=\bm{k}_1-\bm{p}_1}
\\ \nonumber
&&\times\left.
\frac{\overline{u}(\bm{p}_1,s_1)\Gamma^{\mu\nu} u(\bm{k}_1,\sigma_1)
\overline{v}(\bm{k}_2,\sigma_2)\Gamma^{\kappa\lambda} v(\bm{p}_2,s_2)}{
(k_1-p_1)^2+m_C^2-\mathrm{i}\epsilon}
\right).
\end{eqnarray}
Here the $u$ and $v$ spinors are normalized such that
\begin{eqnarray}
\sum_s u(\bm{p},s)\overline{u}(\bm{p},s)&=&m_e-\gamma\cdot p,
\\
\sum_s v(\bm{p},s)\overline{v}(\bm{p},s)&=&-\,m_e-\gamma\cdot p.
\end{eqnarray}

\section{Constraints on antisymmetric tensors from Bhabha scattering}
\label{sec:results}

Limits on deviations from Standard Model Bhabha scattering are reported in 
Refs.~\cite{ref:tasso,ref:tasso2,ref:pluto,ref:mac,ref:topaz,ref:opal} for
energies $14\,\mathrm{GeV}\le\sqrt{s}\le 136.23\,\mathrm{GeV}$. 
Since $m_C>1$ TeV, the correction (\ref{eq:dsigma1}) to Bhabha scattering
depends only on the product $M_e m_C$ up to corrections of less than $2\%$ 
in the energy range considered here. Furthermore, due to $m_e\ll\sqrt{s}$ the
contribution from the dipole coupling depends only on $a_m^2+a_e^2$. 
Therefore we report limits on $\tilde{M}_e m_C$ where $\tilde{M}_e = M_e/\sqrt{a_m^2+a_e^2}$.

 We used 11 data sets published in  
Refs.~\cite{ref:tassod,ref:tasso2d,ref:plutod,ref:macd,ref:topazd,ref:opald} and tabulated 
the statistical/systematic errors as a fraction of the measured cross section. 
To obtain lower bounds on $\tilde{M}_em_C$, we find the required value of the product such that 
the analytically obtained correction ratio drops below the reported error fractions. 
Since measurements of Bhabha scattering cross sections have never detected deviations from the 
Standard Model, this is tantamount to forcing the KR corrections to be smaller than the error bars 
in experimental data. Our results are tabulated in Table \ref{tab:1}.

\begin{table}[ht]
\begin{center}
\begin{tabular}{r|l|l}\hline
Ref. & $\sqrt{s}$ &  Bound on $\tilde{M}_e m_C$ \\ \hline
     & GeV & $10^4\,\mathrm{GeV}^2$ \\ \hline
\cite{ref:tassod} & 34.5 & 5.4 \\ \hline
\cite{ref:tasso2d} & 14 & 1.5 \\ \hline
\cite{ref:tasso2d} & 22 & 2.0 \\ \hline
\cite{ref:tasso2d} & 34.8 & 6.1 \\ \hline
\cite{ref:tasso2d} & 38.3 & 3.5 \\ \hline
\cite{ref:tasso2d} & 43.6 & 5.3 \\ \hline
\cite{ref:plutod} & 34.7 & 5.0 \\ \hline
\cite{ref:macd} & 29 & 6.0 \\ \hline
\cite{ref:topazd} & 52 & 3.4 \\ \hline
\cite{ref:opald} & 130.26 & 7.1 \\ \hline
\cite{ref:opald} & 136.23 & 7.1 \\ \hline
\end{tabular}
\caption{Lower bounds on $\tilde{M}_e m_C$.}
\label{tab:1}
\end{center}
\end{table}

Refs.~\cite{ref:tassod,ref:tasso2d,ref:plutod,ref:macd} include both statistical
and systematic uncertainties, whereas Refs.~\cite{ref:topazd,ref:opald} report
statistical uncertainties on HEPData. However, the discussion in \cite{ref:opal}
shows that the systematic uncertainties are much smaller than the statistical
uncertainties for the OPAL measurements. 

The strongest bound
turns out to be $\tilde{M}_em_C\ge 7.1\times 10^4\, \mathrm{GeV}^2$ from the 130.26 GeV 
and 136.23 GeV measurements of OPAL \cite{ref:opal}. 
If instead we only consider datasets where both statistical 
and systematic errors for cross section measurements were added in quadrature
and reported, the strongest bound for $\tilde{M}_em_C$ 
is $\tilde{M}_em_C\ge 6.1\times 10^4\, \mathrm{GeV}^2$ from 
the 34.8 GeV measurements of TASSO \cite{ref:tasso2}.

\section{Conclusions}
\label{sec:conc}

String theory is still the most compelling framework for particle and gravitational 
physics beyond the Standard Model. 
As such, it behooves us to seek out all possible avenues to experimental tests of
string theory, and the existence of fundamental antisymmetric tensor fields is a 
unique prediction of string theory that should be tested at future facilities.

The clean initial states at lepton colliders will help to push the precision frontier 
in particle physics, and as a first study into signatures of
antisymmetric tensors in M\o{}ller or Bhabha scattering at colliders, we report
constraints from published data of previous experiments. We assume $m_C>1$ TeV from
the absence of BSM resonances at the LHC and find the strongest 
constraint $\tilde{M}_e m_C\ge 7.1\times 10^4\, \mathrm{GeV}^2$ from data published
by OPAL at $\sqrt{s}=130.26$ GeV and $\sqrt{s}=136.23$ GeV.


\acknowledgments
We acknowledge support through the Shastri Indo-Canadian Institute
and the MITACS Globalink program. We also thank NSERC for support through a 
subatomic physics discovery grant.\\

\begin{center}
\textbf{Data Availability Statement}\\[2mm]
\end{center}
This manuscript has no associated data or the data will not be deposited.
[Authors' comment: We did not generate new experimental data as part of this 
investigation. Data used in this paper are publicly available in 
Refs.~\cite{ref:tassod,ref:tasso2d,ref:plutod,ref:macd,ref:topazd,ref:opald}.
We thank the Collaborations for having made their data available on HEPData.]


\begin{thebibliography}{88}
\bibitem{note1}
As noted in \cite{ref:adrd}, the Kalb-Ramond framework for
string interactions can be generalized to the case of $N$ open string species with
boundary charges $g_{s,I}$, $1\le I\le N$, and corresponding boundary gauge fields $\mathcal{B}_{I,\mu}$.
The KR gauge invariant antisymmetric tensor field in this case is
\begin{equation}
C_{\mu\nu}=\mathcal{C}_{\mu\nu}-\frac{1}{\mu_s}\sum_I g_{s,I}\mathcal{B}_{I,\mu\nu}.
\end{equation}
A single $\mathcal{B}_\mu$ can be gauged away through a KR gauge transformation, but $N-1$
boundary gauge fields will always remain.
\bibitem{note2}
Since $\sqrt{s}\ll m_C$ for the experiments considered here, the
parameter that makes the correction (\ref{eq:dsigma1}) small compared to Standard
Model cross sections is actually $\alpha_S v_h^2 s/M_e^2 m_C^2$.
\end{thebibliography}
\end{document}